# Revisiting the Amplitude of Solar Cycle 9: The Case of Sunspot Observations by W.C. Bond

V.M.S. Carrasco[1,2] • M.C. Gallego[1,2] • R. Arlt[3] • J.M. Vaquero[2,4]

[1] Departamento de Física, Universidad de Extremadura, 06006 Badajoz, Spain [e-mail: vmscarrasco@unex.es]

[2] Instituto Universitario de Investigación del Agua, Cambio Climático y Sostenibilidad (IACYS), Universidad de Extremadura, 06006 Badajoz, Spain

[3] Leibniz Institute for Astrophysics Potsdam, An der Sternwarte 16, 14482 Potsdam, Germany

[4] Departamento de Física, Universidad de Extremadura, 06800 Mérida, Spain

**Abstract:** William Cranch Bond, director of the Harvard College Observatory in mid-19th century, carried out detailed sunspot observations during the period 1847–1849. We highlight Bond was the observer with the highest daily number of sunspot groups observed in Solar Cycle 9 recording 18 groups on 26 December 1848 according to the current sunspot group database. However, we have detected significant mistakes in these counts due to the use of sunspot position tables instead of solar drawings. Therefore, we have revisited the sunspot observations made by Bond, establishing a new group counting. Our new counts of the sunspot groups from Bond's drawings indicate that solar activity was previously overestimated. Moreover, after this new counting, Bond would not be the astronomer who recorded the highest daily group number for Solar Cycle 9 but Schmidt with 16 groups on 14 February 1849. We have also indicated the new highest annual group numbers recorded by any observer for the period 1847–1849 in order to correct those values applied in the "brightest star" method, which is used as a rough indicator of the solar activity level. Furthermore, a comparison between Bond's sunspot records and the sunspot observations made by Schwabe and Wolf is shown. We conclude that the statistics of Wolf and Bond are similar regarding to the group count. Additionally, Schwabe was able to observe smaller groups than Bond.

## 1. Introduction

The sunspot record carried out by astronomers since December 1610, when Harriot recorded the first sunspot observation by telescope, is considered as the longest ongoing systematic direct observation set in the world (Owens, 2013; Arlt and Vaquero, 2020).

The sunspot number indices are calculated from the number of sunspots recorded during the telescopic era (Vaquero, 2007; Clette *et al.*, 2014; Usoskin, 2017). Several works have detected problems in the methodology used to calculate these indices (Svalgaard, 2010; Cliver and Ling, 2016; Lockwood, Owens, and Barnard, 2016). For that reason, new sunspot number series have been published to solve it (Lockwood *et al.*, 2016; Clette and Lefèvre, 2016; Svalgaard and Schatten, 2016; Usoskin *et al.*, 2016; Chatzistergos *et al.*, 2017; Willamo, Usoskin, and Kovaltsov, 2017). Furthermore, other works (Carrasco *et al.*, 2015; Neuhäuser and Neuhäuser, 2016; Carrasco *et al.*, 2019) identified problematic observations in the sunspot observation database from which these indices are built. A new revised collection of sunspot group numbers was published (Vaquero *et al.*, 2016, hereafter V16) including corrections for those records and incorporating new data. Nevertheless, Muñoz-Jaramillo and Vaquero (2019) have shown significant discrepancies between the new series, mainly in the historical part, and the weaknesses of V16. Therefore, we must continue analyzing the methodologies used to calculate these indices, incorporate new information to the sunspot observation database, and analyze the problematic records available.

Regarding the solar activity during the 19th century, we can highlight the contributions of Heinrich Schwabe and Rudolf Wolf. Schwabe estimated for the first time the solar cycle periodicity establishing it around 10 years and Wolf defined the relative sunspot number, which is the basis for the currently used International Sunspot Number. Wolf became the first director of the Swiss Federal Observatory (*Eidgenössische Sternwarte*) in 1864 and laid the foundation for a century-long sunspot observing program. Two significant works on these astronomers have been recently published: (i) Arlt *et al.* (2013) calculated the sunspot positions and sizes from solar drawings made by Schwabe for the period 1825–1867, and (ii) Friedli (2016) analyzed the Wolf's original handwritten sourcebooks archived at the library of the University of Zürich. More recently, several works have analyzed the sunspot records made by astronomers of that time, shedding light on solar activity during the first half of the 19th century (mainly during the Dalton Minimum, a period of reduced solar activity that occurred during the first third of the 19th century). For example, Denig and McVaugh (2017) published a set of sunspot drawings made by Jonathan Fisher in 1816 and 1817 and Carrasco *et al.* (2018) analyzed Hallaschka's sunspot observations made in 1814 and 1816, both datasets around the maximum of Solar Cycle 6. Moreover, Hayakawa *et al.* (2020) have studied the sunspot drawings recorded by Derfflinger for the period 1802–1824 providing the sunspot positions and the number of sunspot groups.

After the Dalton Minimum, William Cranch Bond carried out very detailed sunspot observations at Harvard College Observatory for the period 1847–1849. In addition to the publication of his sunspot drawings, Bond also provided information on heliographic positions of some sunspots. The objective of this work is to analyze these sunspot drawings made by Bond during Solar Cycle 9 and to provide a reliable series of the number of sunspot groups for this observer. We indicate some biographical and professional data about Bond in Section 2. In Section 3, we analyze and discuss the sunspot observations recorded by him. Section 4 is devoted to compare the sunspot observations made by Bond, Schwabe, and Wolf. Finally, we show in Section 5 the main conclusion of this work.

**2. Some Notes About William Cranch Bond**

William Cranch Bond (1789–1859) was an American astronomer considered like one of the first important contributors to the early history of the astronomy in the United States (Lockyer, 1897; Stephens, 1990). His attention to astronomy was motivated by the total solar eclipse of 1806 (Holden, 1897). Bond became the first director of Harvard College Observatory in 1839 and carried out different studies in astronomy (Baker, 1895; Hirshfeld, 2015; Bennet, 2020). For example, he discovered the Great Comet of 1811, independently of Honoré Flaugergues. Bond and his son and successor, George Phillips Bond, discovered the Saturn's moon Hyperion and were the first ones to observe the innermost ring of Saturn in 1850. In addition, the experiments in celestial body photography, together with John Adams Whipple, were also noticeable. In fact, in 1850, they provided the first image of any star different to the Sun by photographing Vega, the brightest star of the Lyra constellation. As recognition to his work, one of the Moon craters was named "Bond" and a particular region on Hyperion (a Saturn's moon discovered by him, his son, and William Lassell) was named as Bond-Lassell Dorsum.

Bond (1871) also recorded sunspots at Harvard College Observatory (42° 22′ 53″ N, 71° 7′ 42″ W). His sunspot observations carried out during the period 1847–1849 were published in the *Annals of the Astronomical Observatory of Harvard*. In this documentary source, we can find one or several sunspot drawings on each observation day (sometimes even four drawings for the same day such as, for example, on 15

September 1847) and sunspot position measurements included in tables (Figure 1, left and central panel). The information about the sunspot positions included in these tables (Figure 1, left panel) represents: (i) number of the plate where the observation can be located, (ii) number of the figure in a given plate, (iii) code assigned to sunspots (only a few specific spots have code), (iv) angle in degrees from the preceding point (indicated in the drawings) towards the northern or upper limb, (v) distance of the sunspot to the center of the drawings in a well-defined scale, and (vi) the solar radius measured in that same scale (its unit was nearly one fourth of an inch). We note that the sunspot positions measured by Bond (1871) were studied by Sánchez-Bajo, Vaquero, and Gallego (2010). Bond observed sunspots by projection with a small equatorial telescope (0.10 m in aperture and 1.5 m of focal length). Additionally, he also provided detailed drawings of remarkable sunspot groups (Figure 1, right panel) sometimes using the great refractor telescope (Bond, 1871, p. 3) installed at Harvard College Observatory (around 0.38 m in aperture and 6 m of focal length), the largest in the United States until 1867 (Bond, 1849, 1856). Plates of the sunspot drawings by Bond (1871) are woodcuts reduced by photography from the 7-inches (around 0.18 m) of the original drawings to 3.5-inches (around 0.09 m) diameters. Bond (1871) indicates that the sunspot drawings were generally made by him but some of them by his son George. However, he did not specify which ones were made by his son.

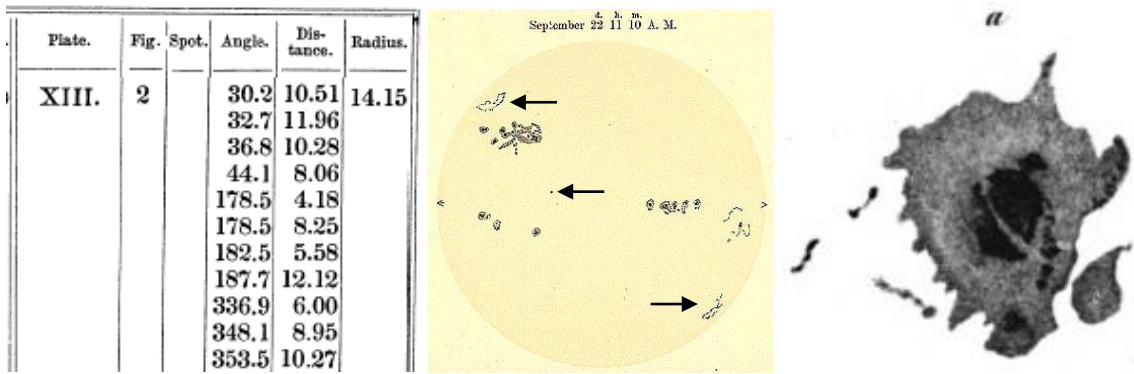

**Figure 1.** Sunspot position measurements (left panel) and drawing (central panel) made by Bond on 22 September 1847. (Right panel) Detailed sunspot (named as "a") recorded by Bond on 16 December 1847 (Bond, 1871). Note that arrows indicate one artefact (not sunspot) and facular regions without sunspots.

**3. Analysis and Discussion of the Bond's Sunspot Observations**

Sunspot observations made by Bond during the period 1847–1849 were made during Solar Cycle 9. According to the sunspot number version 2 (http://sidc.be/silso), this cycle covers the period between 1843 and 1855, with the maximum amplitude in 1848 and it is the solar cycle with the third highest maximum amplitude for the 19th century (only behind Solar Cycle 8 and 11). Figure 2 (top panel) depicts raw daily sunspot group counts included in V16 for the period. Bond's records are represented in red color and those made by the remaining observers in gray color. A great observational coverage is available for this solar cycle, around 95 % regarding all days of the period 1843–1855. In this solar cycle, 20 astronomers recorded sunspot observations. The most active sunspot observers were Schwabe (4113 observation days), Shea (2751), Wolf (1760), and Schmidt (1194).

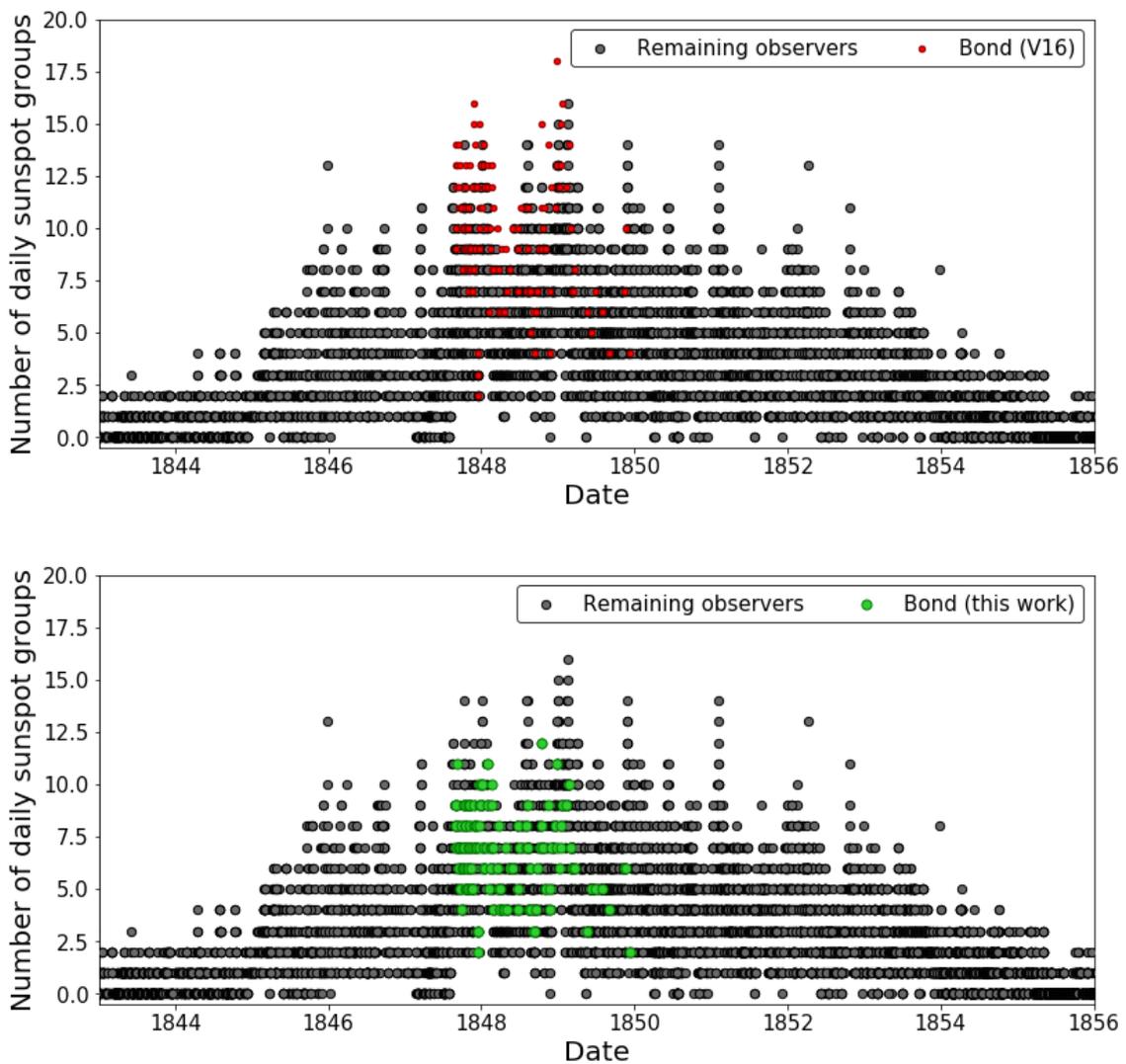

**Figure 2.** (Top panel) Raw daily number of sunspot groups included in V16 from 1843 to 1855. Sunspot observations recorded by Bond are represented in red color and those by the remaining observers in gray color. (Lower panel) Daily number of sunspot groups recorded by Bond according to this work (green color) and the ones included in V16 excluding Bond's counting (gray color) for the period 1843–1855.

Bond, according to V16, carried out 137 sunspot observations from 30 August 1847 to 11 December 1849 (Figure 2, top panel). Thus, Bond is the seventh most active sunspot observers in Solar Cycle 9 regarding the number of records. That record amount implies an observational coverage of 16.4 % for that period. Furthermore, we highlight that Bond recorded the highest daily number of groups observed for Solar Cycle 9. This maximum would correspond to 26 December 1848 when Bond observed 18 groups according to V16.

Hoyt and Schatten (1998) analyzed the sunspot observations made by Bond (1871) and included in their database the number of groups recorded by him. This same group count was incorporated to V16 for Bond. We note that Wolf (1879) mentioned these sunspot observations but he did not provide a sunspot count. We have detected a significant problem in the group counting made by Hoyt and Schatten (1998). These authors considered that the daily number of sunspot groups observed by Bond was equal to the number shown for each day in the tables. However, Bond (1871) actually does not show in those tables the position measurements for all the groups observed each day but only for the most important sunspots. Bond (1871, p. 4) pointed out: "The positions contained in this table are in most cases those of single spots of importance, either from their magnitude or situation upon the limits of their groups; but the positions of the mean of the spots of a group is sometimes given". The number of table entries is therefore smaller than the total number of sunspots on the solar disk, but larger than the number of groups. Thus, the number of sunspot groups recorded by Bond included in V16 is significantly overestimated.

This problem can be clearly seen by comparing the number of sunspot groups recorded in the drawings and the number of sunspot position measurements included in the tables. For example, on 22 September 1847, we can see that the number of sunspot positions measured by Bond was 11 (Figure 1, left panel), which is the number of sunspot groups assigned by Hoyt and Schatten (1998). However, we can count 5 groups in the drawing made by Bond (Figure 1, central panel). Note that there are some artefacts in the drawing, the most striking one located next to the center of the solar disc, and there are no sunspots in two of the three facular regions. These facts are indicated by arrows in the sunspot drawing (Figure 1, central panel).

We may also have a look at the drawing made by Schwabe on 26 December 1848 when V16 lists 18 groups for Bond. Figure 3 shows a comparison of the drawings of the two observers. Note that, using the group code by Schwabe, we have identified the sunspot groups recorded by Bond. Schwabe assigned 11 groups but one can see that at least group 314 consists of two unrelated spots. Group 317 is extremely large and probably plotted in an exaggerated way by Schwabe, but does not appear to consist of more than one group. This is supported by the evolution of the group on the days before and after. Upon redefining the Schwabe groups, Senthamizh Pavai *et al*. (2015) have split group 314 in two, but left groups 315 and 317 intact, since there was no obvious reason to split them. Bond's drawing is very similar to Schwabe's and shows 11 groups. We have also considered in Bond's drawing group 317 as one group. We note that the main difference between both drawings is that Bond recorded for groups 314 and 315 one sunspot on each one and not as Schwabe.

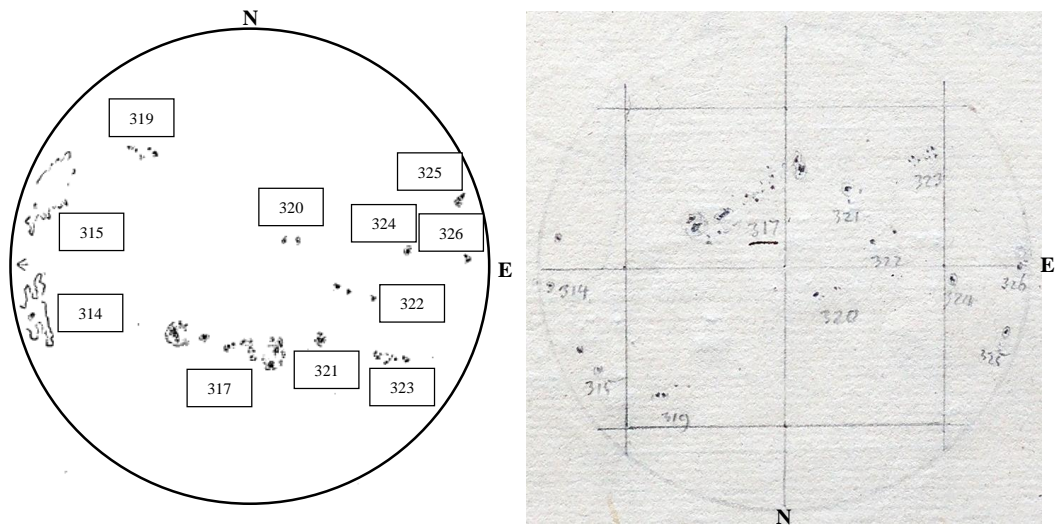

**Figure 3.** Sunspot drawings made by Bond (left panel) and Schwabe (right panel) on 26 December 1848. Note that we have identified the sunspot groups recorded by Bond with those by Schwabe and they have been tagged using the numbers assigned by Schwabe, in addition to the north N and east E in both drawing sets (sources: Bond, 1871; RAS MS Schwabe, courtesy of the Royal Astronomical Society).

We have carried out a new counting of the sunspot groups recorded by Bond (1871) from his drawings. For this purpose, we have followed the modern classifications of sunspot groups (McIntosh, 1990). Figure 4 represents the new count of the daily group number performed in this work (green color) and that one according to Bond's sunspot records included in V16 (red color). This new group count is publicly available on the website of the Historical Archive of Sunspot Observations (haso.unex.es). We can see that the solar activity level calculated from Bond's records included in V16 is significantly greater than that found in this work. Regarding the daily group average for the whole observational period of Bond, we have obtained a value equal to 6.64 while it is 9.65 if it is calculated from V16, i.e. our calculation is a third lower, approximately. We highlight that values of the daily group number obtained in this work are always equal or lower than those included in V16 taking into account the observations made in the same dates. In particular, for the same dates, around 7% of daily group numbers are equal in both counts and the remaining are lower in our group count. The greatest difference in the group counts between both is of 8 groups. This value was found on 15 September (6 groups according to this work versus 14 groups to V16) and 28 November 1847 (8 versus 16) 1847, and 11 January 1848 (6 versus 14). Furthermore, after this new count (Figure 2, bottom panel), the highest daily number of sunspot groups for Solar Cycle 9 is not 18 as recorded by Bond on 26 December 1848, but 16 as recorded by Schmidt on 14 February 1849 (Vaquero *et al.*, 2016). There are five observations assigned to Bond in V16 not included in Bond (1871): 14 September, 30 October, 21 and 23 November 1847, and 3 October 1848. However, we note that Hoyt and Schatten (1998) found sunspot observations made by Bond in Peters' papers at Hamilton College Archive which have not been consulted in this work. We conclude the sunspot counting regarding Bond's data included in V16 should be corrected according to this new analysis.

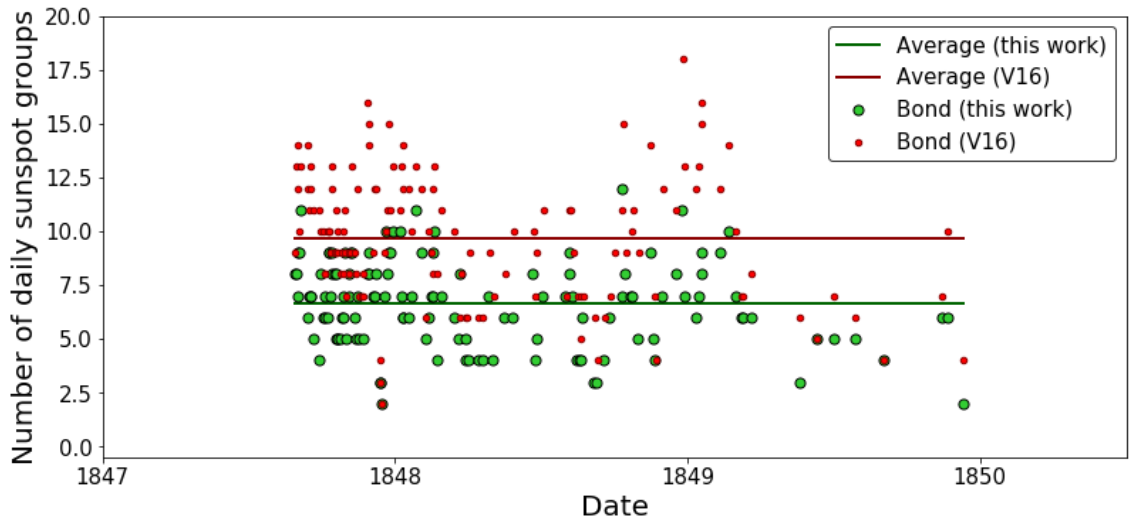

**Figure 4.** Daily sunspot group numbers recorded by Bond for the period 1847–1849 according to this work (green color) and V16 (red color). Horizontal lines depict the daily average.

The sunspot count revision carried out in this work has a high impact in some methodologies proposed to estimate the solar activity level. The "brightest star" method was applied by Svalgaard and Schatten (2016) in order to obtain a rough indicator of solar activity level from the highest group number recorded each year by any observer. It is based on the work of Hubble (1929) to show the relationship between the velocity and distance of galaxies. This methodology, therefore, is very sensitive to the maximum values for each solar cycle. According to V16, the highest values for the period 1847–1849 regarding the daily group number were recorded by: (i) Bond in 1847 (16 groups), (ii) Bond in 1848 (18 groups), and (iii) Bond and Schmidt in 1849 (16 groups). Taking into account the new group count presented in this work, the highest group number for 1847, 1848, and 1849 are: (i) 14 groups recorded by Gerling on 11 October 1847, (ii) 14 groups recorded by Schmidt on 4 January 1848 and 6 August 1848 and by Schwabe on 7 August 1848, and (iii) 16 groups recorded by Schmidt on 14 February 1849, respectively. Thus, this implies a decrease in the highest group number around 10% for 1847 and 20% for 1848. Instead, the value for 1849 would not undergo changes. Therefore, the maximum value for this solar cycle would reduce around 10% since it decreases from 18 to 16 groups. The analysis carried out in this work shows the importance of a good reconstruction of solar activity from good quality data.

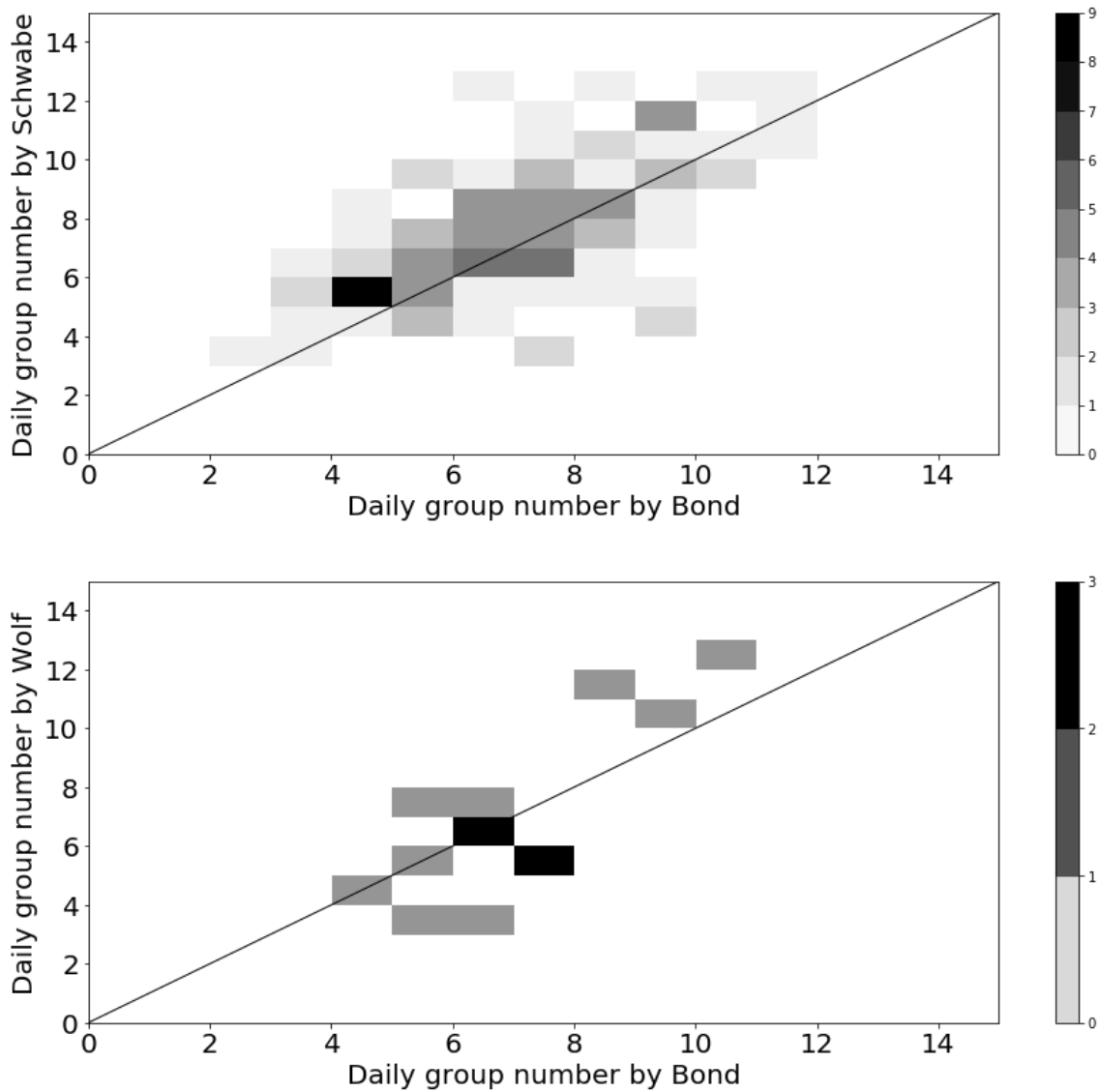

**Figure 5.** Daily number of sunspot groups recorded by Bond (this work) versus Schwabe (top panel) and Wolf (bottom panel). Only coincident observation days have been taken into account in each case. The diagonal line depicts the line of slope equal to 1.

## 4. A Comparison between Bond, Schwabe, and Wolf

We have carried out a comparison between the sunspot observations recorded by Bond and the two main sunspot observers of that time: Schwabe and Wolf. From 30 August 1847 to 11 December 1849, Schwabe recorded 692 observation days, i.e. an observational coverage around 83%. This amount is clearly higher than that obtained for Bond (16.4%). Bond and Schwabe registered 107 sunspot observation days in the same period. Figure 5 (top panel) shows a comparison between the sunspot observations made by both observers in the same observation days. We can see that, in days when

Bond recorded less than 7 groups, Schwabe recorded a greater group number in more observation days than Bond. In this particular case, Schwabe recorded more groups than Bond in 36 observation days, Bond observed more groups than Schwabe in 5 observation days, and the two observers registered the same number of groups in 11 observation days. Regarding the observation days when Bond recorded 7 or more groups, Bond recorded more groups than Schwabe in 21 observation days, Schwabe also observed more groups than Bond in 21 observation days, and the two astronomers recorded the same number of groups in 13 observation days. We note that the greatest differences between the number of groups recorded by both observers in the same date were when Schwabe observed 6 groups more than Bond (9 October 1847 and 7 August 1848) and Bond recorded 5 groups more than Schwabe (3 September 1847 and 8 August 1848). The daily group number average calculated from Schwabe's records is slightly greater than that from Bond. The group average according to Bond is 6.5 groups per day whereas it is 7.1 for Schwabe. Taking into account days when Bond recorded less than 7 groups, the group average for Bond's records is 4.8 and it is 6.0 for Schwabe. When Bond observed 7 or more groups, the group average calculated from both datasets is 8.2 in both cases. These facts suggest that Schwabe was able to record smaller groups than Bond. This can be seen in Figure 6. The smallest groups (coded 247 and 250) recorded by Schwabe (bottom panel) on 13 and 14 December 1847 were not recorded by Bond (top panel).

The coincident observation interval between Wolf and Bond begins from 9 September 1848, when Wolf started his systematic sunspot observations, to 11 December 1849. In this interval, Wolf recorded 279 observation days, that correspond to an observational coverage equal to 60.8%, significantly greater than the around 7% obtained by Bond for the same period. Figure 5 (bottom panel) compares 13 sunspot observations made by these observers in the same dates. We find that they are very similar. Bond recorded more groups than Wolf in 4 days and Wolf observed more groups than Bond in 5 days. Furthermore, the daily group average calculated from each dataset is the same (6.5). We can see that, considering the coincident observation days when Bond recorded less than 7 groups, the group average per day obtained from Bond's records is 5.4 while it is 5.1 according to Wolf. Instead, when Bond recorded 7 or more groups, the group average according to Wolf is 8.6 and it is 8.2 from Bond's records. Moreover, in this particular case, Wolf recorded in 3 days more groups than Bond while Bond observed more groups than Wolf in 2 days. The greatest differences in the number of groups were on 18 January 1849 (Wolf recorded 11 groups and Bond 8 groups) and 9 March 1849 (Bond observed 6 groups and Wolf 3 groups). This similarity in the observations suggests that differences in the sunspot group count recorded by Bond and Wolf can be in the methodology to define groups. Furthermore, we highlight that Bond, Schwabe, and Wolf made sunspot observations in 11 coincident days. The daily group average calculated according to the sunspot records made by each observer is similar: 6.4, 6.6, and 6.7 for Bond, Schwabe, and Wolf, respectively. Taking into account the sunspot records made by these three observers, Wolf recorded the maximum number of groups in 5 coincident observation days, while Schwabe and Bond did it in 4 and 3 days, respectively.

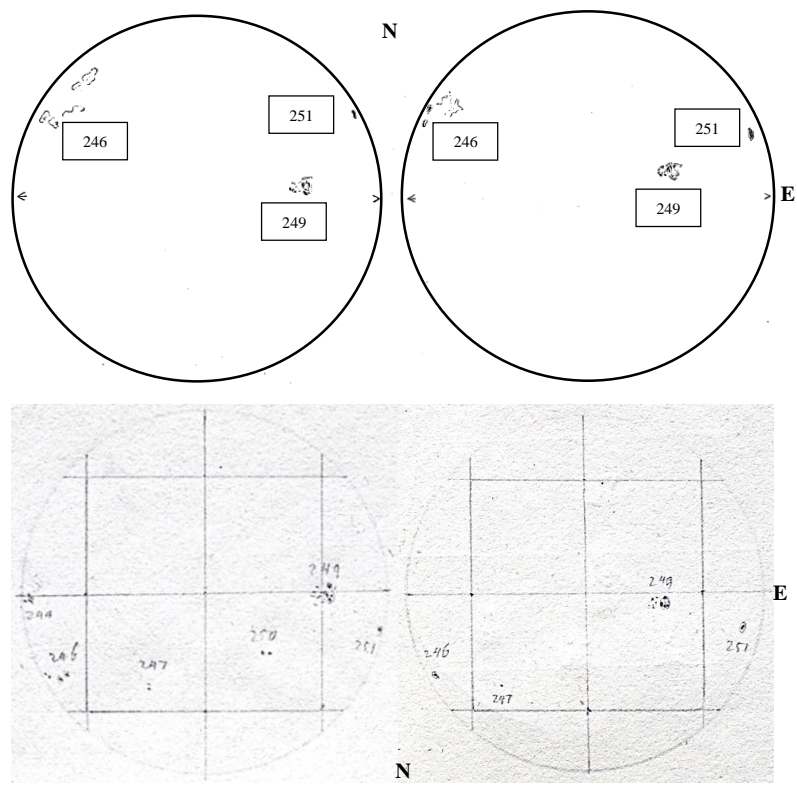

**Figure 6.** Sunspot drawings made by Bond (upper panel) and Schwabe (lower panel) on 13 (left panel) and 14 (right panel) December 1847. Note that we have identified the sunspot groups recorded by Bond with those by Schwabe and they have been tagged using the numbers assigned by Schwabe, in addition to the north N and east E in both drawing sets (sources: Bond, 1871; RAS MS Schwabe, courtesy of the Royal Astronomical Society).

## 5. Conclusions

William Cranch Bond made sunspot observations during the period 1847–1849, i.e. around the maximum of Solar Cycle 9. Bond published his sunspot observations in the *Annals of the Astronomical Observatory of Harvard* including sunspot drawings and tables with sunspot position measurements. These sunspot records were previously analyzed by Hoyt and Schatten (1998) and Vaquero *et al*. (2016) incorporated the group count made by Hoyt and Schatten (1998) to the current sunspot group database. We have identified significant mistakes in that group count because Hoyt and Schatten (1998) determined the group number recorded by Bond from the tables where sunspot positions are listed. However, Bond mainly represented in those tables the sunspot positions of the most important single sunspots and not group positions.

We have carried out an analysis of the sunspot observations made by Bond. In this work, we present a new counting according to his sunspot drawings. The solar activity level calculated from the new group counting is one third lower than that obtained from V16 for the same dataset. Moreover, after this new group count, the highest daily group number for Solar Cycle 9 was actually not recorded by Bond on 26 December 1848 (18 groups) but by Schmidt on 14 February 1849 (16 groups). Applying the new group count presented in this work to the "brightest star" method applied by Svalgaard and Schatten (2016) as a rough indicator of solar activity level, the year 1849 would not be affected but it would imply a decrease of around 10% for 1847 and 20% for 1848 in the highest group number recorded each year. Furthermore, comparing Bond's sunspot records with sunspot observations made by Wolf and Schwabe, we can see that the ones by Bond and Wolf are similar in terms of group count. On the other hand, Schwabe recorded more groups than Bond because he was able to observe smaller ones.

This work brings out the need to continue analyzing the historical sunspot records as well as to incorporate new information in order to obtain a sunspot group database with good quality observations. Only in this way we can calculate reliable solar activity indices for past centuries.

**Acknowledgements**

This research was supported by the Economy and Infrastructure Counselling of the Junta of Extremadura through project IB16127 and grants GR18081 (co-financed by the European Regional Development Fund) and by the Ministerio de Economía y Competitividad of the Spanish Government (CGL2017-87917-P). The authors have benefited from the participation in the ISSI workshops led by M.J. Owens and F. Clette on the calibration of the sunspot number.

**Disclosure of Potential Conflicts of Interest and Ethical Statement**

The authors declare that they have no conflicts of interest. All authors contributed to the study conception and design. The first draft of the manuscript was written by V.M.S. Carrasco and all authors read and approved the final manuscript.